\documentclass[twocolumn,aps,prb,showpacs]{revtex4}

\usepackage{epsfig}

\begin{document}

\title {Quantum interference between non-magnetic impurities
in $d_{x^2-y^2}$-wave superconductors}
\author{Dirk K.~Morr and Nikolaos A. Stavropoulos}
\affiliation{Department of Physics, University of Illinois at
Chicago, Chicago, IL 60607}
\date{\today}
\begin{abstract}
We study quantum interference of electronic waves that are
scattered by multiple non-magnetic impurities in a
$d_{x^2-y^2}$-wave superconductor. We show that the number of
resonance states in the density-of-states (DOS), as well as their
frequency and spatial dependence change significantly as the
distance between the impurities or their orientation relative to
the crystal lattice is varied. Since the latter effect arises from
the momentum dependence of the superconducting gap, we argue that
quantum interference is a novel tool to identify the symmetry of
unconventional superconductors.
\end{abstract}


\maketitle

Over the last few years, the study of impurities in unconventional
superconductors has attracted considerable theoretical
\cite{Bal95,Fla97,Haas00,Pol01} and experimental
\cite{Pan00,Hud01,Der02} attention. In particular, a series of
groundbreaking scanning tunneling microscopy (STM) experiments has
provided detailed information on the density-of-states (DOS) near
single non-magnetic \cite{Pan00} and magnetic \cite{Hud01}
impurities in Bi$_2$Sr$_2$CaCu$_2$O$_{8+\delta}$, a
high-temperature superconductor (HTSC). Of particular interest is
the experimentally observed emergence of resonance states near the
impurities. Several theoretical scenarios have been proposed that
ascribe the origin of these resonances to electronic scattering
off classical impurities with magnetic and non-magnetic scattering
potentials \cite{Bal95,Fla97,Haas00} or to the Kondo-screening of
a local spin polarization that is induced by non-magnetic
impurities \cite{Pol01}. At the same time, a different line of
beautiful experiments studied quantum interference of electronic
waves that are scattered by multiple impurities. In particular,
Manoharan {\it et al.}~\cite{Man00} demonstrated that quantum
interference in a corral of magnetic impurities arranged on the
surface of a metallic host leads to the focusing of electronic
waves into a quantum image; a result that has recently been
addressed in several theoretical studies \cite{QCtheory}. Similar
quantum interference experiments using unconventional
superconductors can be expected in the future. First evidence for
quantum interference in the HTSC was recently reported by Derro
{\it et al.}~\cite{Der02} in the one-dimensional chains of
YBa$_2$Cu$_3$O$_{6+x}$.

Motivated by the experimental progress in this field, we present
in this article a theoretical model which combines the study of
impurities in unconventional superconductors with that of quantum
interference effects. In particular, generalizing the formalism
presented in \cite{Bal95,Fla97,Morr01}, we investigate the
electronic structure in the vicinity of two non-magnetic
impurities in a $d_{x^2-y^2}$-wave superconductor. We show that
quantum interference due to the presence of a second impurity
dramatically changes the DOS obtained near a single impurity. In
particular, we demonstrate that the number of resonance states in
the DOS, as well as their frequency and spatial dependence change
significantly as the distance between the impurities or their
orientation relative to the crystal lattice is varied. Since the
latter effect arises from the momentum dependence of the
superconducting gap, we argue that quantum interference is a novel
tool to identify the symmetry of unconventional superconductors.
This result might be of particular importance for Sr$_2$RuO$_4$,
an unconventional superconductor whose symmetry is still a topic
of controversy \cite{Maeno01}. While the study of quantum
interference is not only of fundamental importance for our
understanding of complex impurity structures, it can also clarify
the origin of the resonances observed in the HTSC. In particular,
we expect that the form of the resonances arising from
Kondo-screening of two magnetic impurities is different from those
discussed below; work is currently under way to verify this
conjecture \cite{Morr02b}.

Starting point for our calculations is the $\hat{T}$-matrix
formalism \cite{Shiba68} which we extended to treat scattering off
multiple impurities \cite{Bal95,Morr01} in a $d_{x^2-y^2}$-wave
superconductor. Quantum interference in $s$-wave superconductors
was recently discussed in Refs.\cite{Fla00,Morr02a}. For
simplicity we restrict our considerations to two non-magnetic
impurities; our formalism, however, allows the study of an
arbitrary, but finite number of impurities. The study of more
complex impurity structures, as well as that of magnetic
impurities, will be the focus of future work \cite{Morr02b}.
Within the Nambu-formalism and for Matsubara frequencies,
$\omega_n$, the electronic Greens function in the presence of $N$
impurities is given by
\begin{eqnarray}
\hat{G}(r,r',\omega_n)&=&\hat{G}_0(r,r',\omega_n) \nonumber \\
& & \hspace{-2cm} +\sum_{i,j=1}^N
\hat{G}_0(r,r_i,\omega_n)\hat{T}(r_i,r_j,\omega_n)\hat{G}_0(r_j,r',\omega_n)
\ , \label{Ghat}
\end{eqnarray}
where the $\hat{T}$-matrix is obtained from the Bethe-Salpeter
equation
\begin{eqnarray}
\hat{T}(r_i,r_j,\omega_n)&=&\hat{V}_{r_i}\delta_{r_i,r_j}
  \nonumber \\
& & \hspace{-1.5cm} +\hat{V}_{r_i}\sum_{l=1}^N
\hat{G}_0(r_i,r_l,\omega_n)\hat{T}(r_l,r_j,\omega_n) \ .
\label{T1}
\end{eqnarray}
For two non-magnetic impurities located at ${\bf r}_i$ ($i=1,2$)
with $\Delta r=|{\bf r}_2 - {\bf r}_1|$,  the scattering matrices
are given by $\hat{V}_{r_i}=U_i \, \tau_3/2$ with $U_{i}$ being
the non-magnetic scattering potential and $\tau$ the
Pauli-matrices in Nambu-space. The Greens function of the
unperturbed (clean) system in momentum space is given by
\begin{equation}
\hat{G}^{-1}_0({\bf k},i\omega_n)=\left[ i\omega_n \tau_0 -
\epsilon_{\bf k} \tau_3 \right] - \Delta_{\bf k} \tau_1 \ .
\label{VG}
\end{equation}
For the electronic excitation spectrum in the normal state we take
a form that is characteristic of an optimally doped HTSC
\cite{JC,Mes99,ZX} and given by
\begin{equation}
\epsilon_{\bf k} = -2t \Big[ \cos(k_x) + \cos(k_y) \Big]
 -4t^\prime \cos(k_x) \cos(k_y) -\mu \ ,
\end{equation}
with $t=300$ meV, $t^\prime/t=-0.4$ as the nearest and
next-nearest neighbor hopping integrals, respectively, and a
chemical potential $\mu/t=-1.18$, corresponding to $14\%$ hole
doping. Moreover, the superconducting gap with
$d_{x^2-y^2}$-symmetry is given by $\Delta_{\bf k}=\Delta_0
\left[\cos k_x-\cos k_y \right] /2 $ with $\Delta_0 =25$ meV
\cite{Mes99}. Our results presented below are qualitatively robust
against changes in the form of $\epsilon_{\bf k}$, or the size of
$\Delta_{\bf 0}$. We obtain the DOS, $N({\bf r},\omega)$, from a
numerical evaluation of Eqs.(\ref{Ghat})-(\ref{VG}) with $N({\bf
r},\omega)=A_{11}+A_{22}$ and $A_{ii}({\bf r},\omega)=-{\rm Im}\,
\hat{G}_{ii}({\bf r},\omega+i\delta)/ \pi$.

We briefly review some important features in the DOS near a single
non-magnetic impurity in a $d_{x^2-y^2}$-wave superconductor
\cite{Bal95,Fla97}. The resulting diagonal ${\hat T}$-matrix
\begin{equation}
{\hat T}_{11,22}=\frac{ \pm U_0}{ 1-U_0 \, G_0(r=0,\pm \omega)}
\label{singleT}
\end{equation}
where the upper (lower) sign applies to ${\hat T}_{11} ({\hat
T}_{22})$ and $G_0=[{\hat G}_0]_{11}$, exhibits a particle-
($\omega_{res}<0$) and hole-like ($\omega_{res}>0$) resonance.
These resonances give rise to sharp peaks in the DOS only in the
unitary limit ($|\omega_{res}|/\Delta_0 \ll 1$) where $U^{-1}_0=
{\rm Re}\, G_0(0,\pm \omega_{res})$.

In the presence of two impurities, the ${\hat T}$-matrix,
Eq.(\ref{T1}), is rather complex. However, in the limit
$F_0(\Delta r,\omega) \ll G_0(\Delta r,\omega)$, where $F_0=[{\hat
G}_0]_{12}$ and for identical impurities with $U_{1,2}=U_0$, the
${\hat T}$-matrix simplifies considerably and is again diagonal.
Defining
\begin{equation}
S_{\pm}(\omega)= \left\{ 1-U_0 \left[ G_0(0,\omega) \pm G_0(\Delta
r , \omega) \right] \right\}^{-1}
\end{equation}
one obtains ($ i \not = j$)
\begin{eqnarray}
{\hat T}_{11,22}(r_i, r_i)&=&U_0 \left[ S_{+}(\pm
\omega)+S_{-}(\pm \omega)\right]/2 \nonumber \\
{\hat T}_{11,22}(r_i,r_j)&=&  \frac{U_0 \, G_0(\Delta r , \pm
\omega)}{ 1-U_0 G_0(0,\pm \omega)} {\hat T}_{11,22}(r_i,r_i)
\label{twoT}
\end{eqnarray}
where the upper (lower) sign applies to ${\hat T}_{11} ({\hat
T}_{22})$. By comparing the ${\hat T}$-matrices in
Eqs.(\ref{singleT}) and (\ref{twoT}), we find that the presence of
a second impurity splits the resonances of the single impurity
case by $U_0 G_0(\Delta r, \omega)$. Note that $G_0(\Delta r,
\omega)$ does not only change with varying $\Delta  r$, but also
with the angle, $\alpha$, between ${\bf r}_2 - {\bf r}_1$ and the
crystal $\hat{x}$-axis, due to the momentum dependence of the
superconducting gap. Consequently, the energy and lifetime of the
resonances depend on $\Delta  r$ and $\alpha$. While all four
$S_{\pm}$-terms in Eq.(\ref{twoT}) can possess resonances, those
that are shifted to higher energies are highly overdamped and give
rise only to oscillations in the DOS without the signature of a
sharp peak.

%
%
\begin{figure}[t]
\epsfig{file=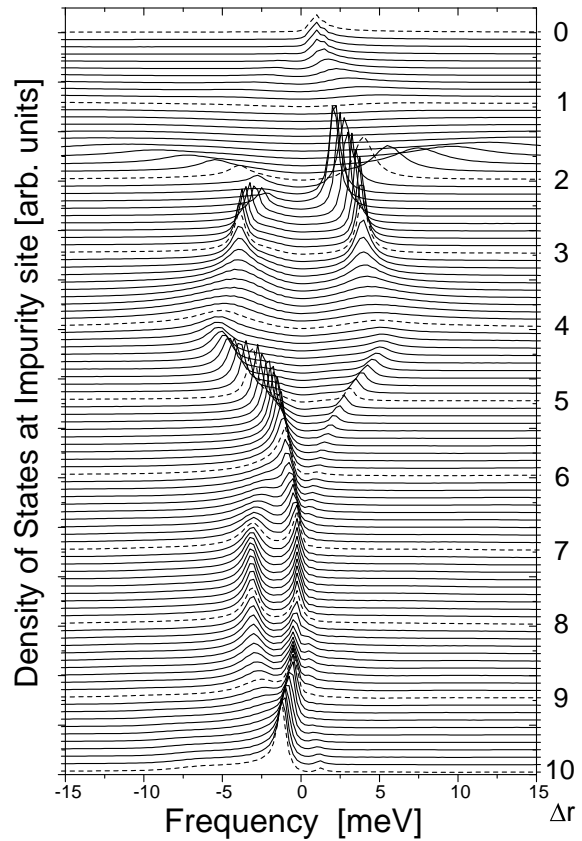,width=7.5cm} \caption{DOS at ${\bf R}=(0,0)$
as a function of $\Delta r$ for two identical impurities with
$U_0=700$ meV located at ${\bf r}_1=(0,0)$ and ${\bf r}_2=(\Delta
r,0)$. The lattice constant is set to $a_0=1$.} \label{U700}
\end{figure}
In what follows, we consider two identical impurities with
scattering potential $U_{1,2}=U_0=700$ meV, corresponding to the
unitary limit, in agreement with Refs.\cite{Bal95,Fla97}. While
the specific form of the DOS changes with $U_0$, its qualitative
features remain unchanged. To study the effects of $\alpha$ and
$\Delta r$ on the DOS separately, we first consider for
definiteness two impurities located along the crystal ${\hat
x}$-axis at ${\bf r}_1=(0,0)$ and ${\bf r}_2=(\Delta  r,0)$ with
$\alpha=0$. In Fig.~\ref{U700}, we present the DOS at ${\bf
R}=(0,0)$, i.e., at one of the impurity sites, as a function of
$\Delta r$. For comparison, we note that for a single impurity
with $U_0=700$ meV, the resonances are located $\omega_{res}=\pm
1.5 $ meV. As $\Delta r$ is varied, the DOS undergoes strong
modifications. In particular, the frequency of the resonances
oscillates and at the same time, their energy width, or lifetime,
changes. For a single impurity, the resonance frequency and width
are correlated, such that as $|\omega_{res}|$ decreases, the
frequency width decreases as well \cite{Bal95,Fla97}. In the case
of two impurities, we find that $|\omega_{res}|$ and the lifetime
of the resonances are not necessarily correlated. For example, the
resonance frequencies for both $\Delta r=2.0$ and $\Delta  r=3.5$
are $\omega_{res}= \pm 4.0$ meV, but the width of the resonances
are considerably larger in the second case. Moreover, for some
values of $\Delta r$, {\it all} resonances are very weak and,
e.g., for $\Delta  r \approx 1$ disappear almost completely. Note,
that even for rather large values of $\Delta  r \approx 10$, the
DOS at ${\bf R}=(0,0)$ is still affected by quantum interference
and thus different from that obtained in the single impurity case.
This result bears important implications for the interpretation of
recent STM experiments \cite{Pan00,Hud01,Der02} since it implies
that the DOS near impurities in the two-dimensional HTSC can only
be described within the single impurity framework if the impurity
concentration is well below 1\%.

\begin{figure}[t]
\epsfig{file=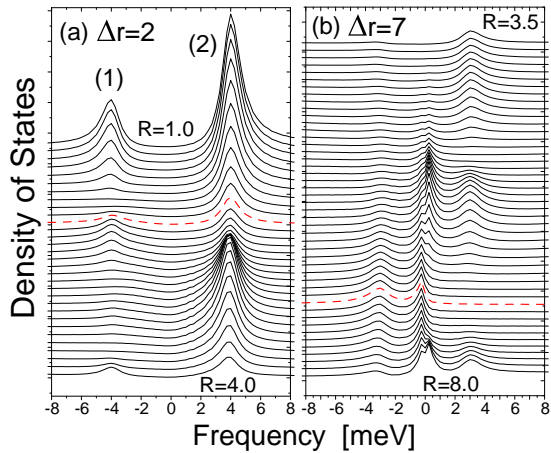,width=7.5cm} \caption{DOS as a function of
spatial position ${\bf R}=(R,0)$ for two impurities with $U_0=700$
meV  located at ${\bf r}_1=(0,0)$ and ${\bf r}_2=(\Delta r ,0)$
with (a) $\Delta r =2$ and (b) $\Delta  r=7$.} \label{dr2dr7}
\end{figure}

An additional important result of Fig.~\ref{U700} is that the
number of observable low-energy resonances changes with $\Delta
r$. In particular, for $\Delta r \leq 6 $, only two sharp
low-energy resonances can be clearly identified. This effect
becomes particularly evident when one considers the spatial
dependence of the DOS for fixed $\Delta r$, as shown in
Fig.~\ref{dr2dr7}. Here, we plot the DOS as a function of ${\bf
R}=(R,0)$ for two impurities located at ${\bf r}_1=(0,0)$ and
${\bf r}_2=(\Delta r,0)$. The uppermost curve corresponds to the
midpoint between the two impurities, the dashed line represents
the DOS at ${\bf r}_2$.
%
%
For $\Delta  r =2 $ (Fig.~\ref{dr2dr7}a), there exist only two
low-energy resonances at $\omega_{res} \pm 4$ meV.  In contrast,
for $\Delta r =7 $ (Fig.~\ref{dr2dr7}b), we find two broader
resonances at $\omega_{res} = \pm 3$ meV, and two sharper
resonances at $\omega_{res} = \pm 0.25$ meV. Note, that the
resonances for $\Delta r =2 $ at $\omega_{res} \pm 4$ meV have a
considerably larger amplitude than those for $\Delta r =7 $ at
$\omega_{res} = \pm 3$ meV. This result again differs from the
single impurity case, where the resonance with the smaller
$|\omega_{res}|$ always possesses a larger amplitude in the DOS.

\begin{figure}[t]
\epsfig{file=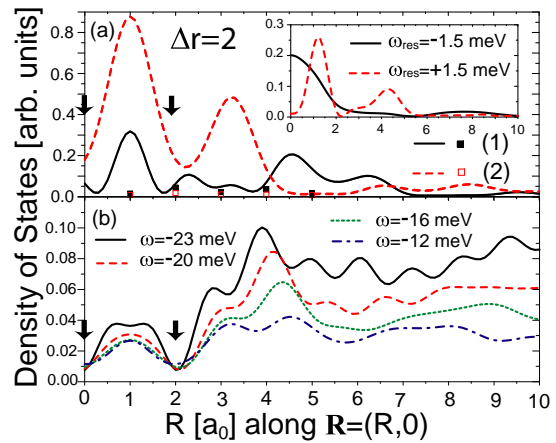,width=7.5cm} \caption{(DOS along ${\bf
R}=(R,0)$ for the same impurity arrangement as in
Fig.~\ref{dr2dr7}a. The positions of the impurities are indicated
by arrows. (a) DOS at $\omega_{res}=\pm 4$ meV. The open and
filled squares present the DOS at $\omega_{res}=\pm 4$ meV along
the lattice diagonal with ${\bf R}=(R,R)$. Inset: DOS along ${\bf
R}=(R,0)$ for a single impurity at its resonance frequency
$\omega_{res}=\pm 1.5$ meV. (b) DOS at frequencies
$|\omega_{res}|<|\omega|<\Delta_0$.} \label{amp_comp}
\end{figure}

While sharp resonances can only be identified for $\omega_{res}
\ll \Delta_0$, oscillations in the DOS exist for basically all
energies $|\omega| \leq \Delta_0$. To study these oscillations in
more detail, we present in Fig.~\ref{amp_comp} the DOS along ${\bf
R}=(R,0)$ for various frequencies and the same impurity
arrangement as in Fig.~\ref{dr2dr7}a; the locations of the
impurities are indicated by arrows.
%
%
The solid and dashed lines in Fig.~\ref{amp_comp}a represent the
DOS at $\omega_{res}=\pm 4$ meV, corresponding to peak (1) and (2)
in Fig.~\ref{dr2dr7}a. The DOS at $|\omega_{res}|$ but along the
lattice diagonal, i.e., for ${\bf R}=(R,R)$, is shown as open and
filled squares. Similar to the single impurity case, the amplitude
of the resonances is much weaker along the direction of the
superconducting gap nodes, than along the anti-nodal direction.
The inset shows the DOS along ${\bf R}=(R,0)$ for a single
impurity with $U_0=700$ meV and resonance energy $\omega_{res} =
\pm 1.5$ meV. A comparison shows that the amplitude of the DOS
oscillations induced by two impurities is much larger than those
induced by a single impurity (the overall scale in the inset is
three times smaller than in the main figure). Moreover, in the two
impurity case, the DOS exhibits significant oscillations at much
larger distances from the impurities than in the single impurity
case. This is particularly evident when comparing the
particle-like resonances, where in the two impurity case, the
amplitude of the oscillations is still large at a distance to the
nearest impurity, $r_n$, of about $r_n \approx 4-5$, while in the
single-impurity case, the oscillations are already substantially
reduced at $r_n \approx 2$. In Fig.~\ref{amp_comp}b we present the
DOS along ${\bf R}=(R,0)$ for several frequencies with
$|\omega_{res}|<|\omega|<\Delta_0$. While there exist no evidence
for a resonance at higher energies, we still find considerable
oscillations in the DOS. As $|\omega|$ decreases, the wave-vectors
of these oscillations decreases, as can clearly be seen from the
shift of the peaks around $R=3$ and $4$. Thus, the DOS
oscillations exhibit a dispersion, similar to the results obtained
in Ref.~\cite{Der02}.

%
%
\begin{figure}[t]
\epsfig{file=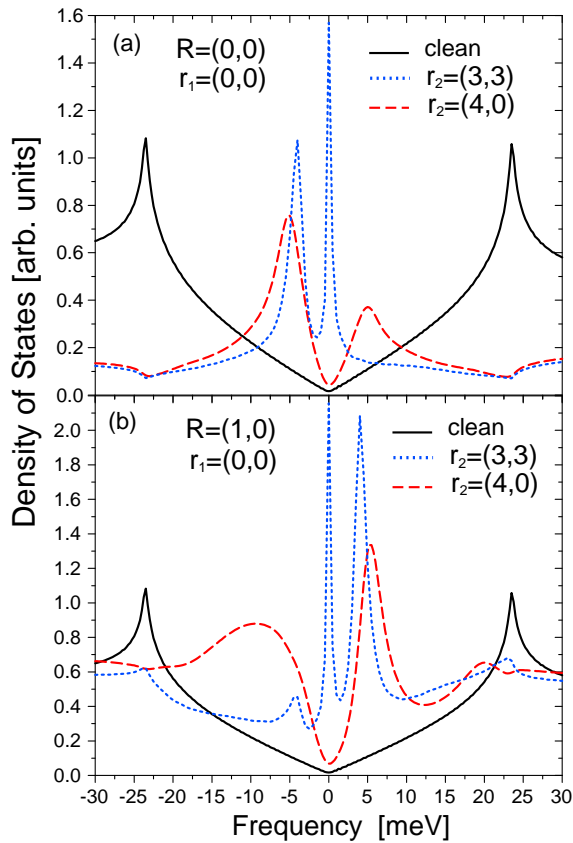,width=7.5cm} \caption{DOS at (a) ${\bf
R}=(0,0)$ and (b) ${\bf R}=(1,0)$ for two impurities with $U=700$
meV. One impurity is located at ${\bf r}_1=(0,0)$, the other one
either at ${\bf r}_2=(3,3)$ (dotted line) or at ${\bf r}_2=(4,0)$
(dashed line).} \label{axis_diag}
\end{figure}
Due to the momentum dependence of the superconducting
$d_{x^2-y^2}$-gap, the DOS changes when the orientation of the two
impurities with respect to the crystal lattice is varied. In
particular, since the gap vanishes along the lattice diagonal, we
expect the largest deviations from the results for $\alpha=0$
shown in Figs.~\ref{U700}-\ref{amp_comp} when the impurities are
located along the lattice diagonal, i.e., for $\alpha=\pi/4$. To
study the changes in the DOS due to variations in $\alpha$ and to
eliminate effects due to a varying $\Delta r$, we chose two
different impurity arrangements which can be realized
experimentally, and possess almost identical values for $\Delta
r$. In the first case, the impurities are located along the
lattice diagonal ($\alpha=\pi/4$) at ${\bf r}_1=(0,0)$ and ${\bf
r}_2=(3,3)$ (dotted line, $\Delta r \approx 4.2$). In the second
arrangement, the impurities are aligned along the crystal ${\hat
x}$-axis ($\alpha=0$) at ${\bf r}_1=(0,0)$ and ${\bf r}_2=(4,0)$
(dashed line, $\Delta r=4$). We verified that the DOS for ${\bf
r}_2=(4.2,0)$, which would yield identical values of $\Delta r$,
but is experimentally not realizable, differs only slightly from
that for ${\bf r}_2=(4,0)$. The DOS at ${\bf R}=(0,0)$ and ${\bf
R}=(1,0)$ is shown in Fig.~\ref{axis_diag}. While the DOS for
$\alpha=\pi/4$ possesses three distinct resonances,  only two
resonance peaks are observable for $\alpha=0$. Moreover, for
$\alpha=0$, the resonance states are located at higher frequencies
and are much broader. The qualitative differences in the DOS
between these two different impurity arrangements can be directly
traced back to the vanishing of $F(\Delta r,\omega)$ for
$\alpha=\pi/4$, and its finite, complex value for $\alpha=0$.
Thus, the symmetry of the superconducting gap is directly
reflected in the changes which the DOS undergoes when the
orientation of the impurities relative to the crystal lattice is
varied. This dependence provides a new tool to identify the
symmetry of unconventional superconductors.

In summary, we studied quantum interference of electronic waves
that are scattered by two non-magnetic impurities in a
$d_{x^2-y^2}$-superconductor. We show that the number of resonance
states in the DOS, as well as their frequency and spatial
dependence changes significantly as the distance between the
impurities or their orientation relative to the crystal lattice is
varied. The latter result provides a novel tool to identify the
symmetry of unconventional superconductors, such as Sr$_2$RuO$_4$,
where the symmetry of the superconducting state is still a topic
of controversy.

We would like to thank J.C. Davis and A. de Lozanne for
stimulating discussions.

\end{document}